\documentclass[useAMS,usenatbib, letterpaper]{mn2e} \usepackage{natbib} 
\usepackage{graphicx} \usepackage{amssymb}

\title[Modelling of SLUGS Gravitational Lenses]{Modelling of the Complex CASSOWARY/SLUGS Gravitational Lenses} \author[B. J. Brewer et al.]{Brendon J. 
Brewer$^1$\thanks{Email: \texttt{brewer@physics.ucsb.edu}}, Geraint F. Lewis$^2$, Vasily 
Belokurov$^3$, Michael J. Irwin$^3$ \and Terry J. Bridges$^{4,5}$, N. Wyn Evans$^3$\\ \\ $^1$Department of 
Physics, University of California, Santa Barbara, CA 93106-9530, USA\\ 
$^2$Sydney Institute for Astronomy, School of 
Physics A28, The University of Sydney, NSW 2006, Australia \\
$^3$Institute of Astronomy, University of Cambridge, Madingley Road, 
Cambridge CB3 0HA, UK\\
$^4$Australian Gemini Office, Anglo-Australian Observatory, PO Box 296, Epping NSW 1710, Australia\\
$^5$Department of Physics, Engineering Physics \& Astronomy, Queen's University, Kingston, Ontario K7L 3N6, Canada}

\begin{document} \date{\today} 
\pagerange{\pageref{firstpage}--\pageref{lastpage}} \pubyear{2009} 
\maketitle \label{firstpage}

\begin{abstract} We present the first high-resolution images of 
 CSWA 31, a gravitational lens system observed as part of the SLUGS (Sloan Lenses 
Unravelled by Gemini Studies) program. These systems exhibit complex image 
structure with the potential to strongly constrain the mass distribution 
of the massive lens galaxies, as well as the complex morphology of the 
sources. In this paper, we describe the strategy used to reconstruct the 
unlensed source profile and the lens galaxy mass profiles. We introduce a prior 
distribution over multi-wavelength sources that is realistic as a 
representation of our knowledge about the surface brightness 
profiles of galaxies and groups of galaxies. To carry out the inference computationally, we use Diffusive Nested 
Sampling, an efficient variant of Nested Sampling that uses Markov Chain Monte Carlo (MCMC) to sample the complex posterior distributions and compute the normalising constant. We demonstrate the efficacy of this approach with 
the reconstruction of the group-group gravitational lens system CSWA 31, finding the source to be composed of five merging spiral galaxies magnified by a factor of 13.
\end{abstract}


\section{Introduction} Gravitational lensing has revealed itself to be a 
powerful probe of not only the distribution of luminous and dark matter within the 
lensing galaxies, but also the small scale structure apparent in sources 
when the images are reconstructed. Initially, with point-like quasar 
sources, only limited information was available to constrain 
reconstructions, but the identification of extended images has led to 
the development of a number of techniques to maximize the 
use of additional information in determining both the lens and source 
properties \citep[for a recent review of strong lensing by galaxies, see][]{2010ARA&A..48...87T}.

The development of large scale surveys has allowed for the automatic 
search and identification of large numbers of gravitational lens systems. A prominent example is SLACS 
(Sloan Lens ACS Survey), which uses HST/ACS to follow-up lens candidates 
identified in the Sloan Digital Sky Survey (SDSS) \citep{slacs}. SLACS 
targets small separation lenses, in which the images are unresolved by 
SDSS, typically through the spectroscopic identification of emission 
lines from high redshift objects within the spectra of lower redshift 
early-type galaxies. Thus, SLACS has been highly successful in using 
lensing to study the properties of typical elliptical galaxies 
\citep{treu}.

Alternatively, the Cambridge And Sloan Surey of Wide ARcs in the skY 
(CASSOWARY) survey 
\citep[][\texttt{http://www.ast.cam.ac.uk/research/cassowary/}]{2009MNRAS.392..104B}, 
explores a different regime of galaxy-galaxy lens systems by searching for multiple 
blue faint companions around massive luminous red galaxies in SDSS. 
Thus, CASSOWARY is suited to finding wide separation lenses with high 
magnification. So far, CASSOWARY has identified 40 systems, with several more 
unconfirmed candidates. These lenses are suited to utilising the unique 
properties of gravitational lensing to provide information about the 
mass profiles of massive galaxies and galaxy groups, and to provide very high resolution images 
of faint high-redshift galaxies. Previous studies of group lenses include \citet{2009A&A...502..445L}.

In this paper we present the first analysis of high resolution 
Gemini/GMOS images of a confirmed CASSOWARY lens system, 
taken as part of the SLUGS (Sloan Lenses Unraveled by Gemini Studies) 
program. The focus of this paper is the reconstruction technique, with 
the details of the extension to other CASSOWARY lenses being presented 
elsewhere (Belokurov et al {\it in prep.}) In Section~\ref{lensing}, we 
describe the approach to gravitational lens modelling that will be adopted for 
the SLUGS systems, discussing the prior source information in 
Section~\ref{priors}, the adopted lens models in Section~\ref{lens}, and 
the implementation of nested sampling in Section~\ref{nested_section}.  The 
second part of the paper, Section~\ref{cswa31}, we will present the 
implementation of this scheme on CSWA 31; a newly confirmed 
gravitational lens and part of the SLUGS sample. The conclusions of 
this study are presented in Section~\ref{conclusions}.

\section{Gravitational Lens Reconstruction}\label{lensing} Gravitational 
lens reconstruction is a classic inference problem 
\citep{2006ApJ...637..608B}: given some data $D$ (i.e. image pixel values) and prior information 
(and/or assumptions), we wish to infer various properties of the lens 
system such as its source parameters $S$, the lens parameters $L$, and 
various other parameters, denoted collectively by $\Phi$; for example, 
$\Phi$ may include parameters describing the light profile of the lens galaxy, and unknown 
noise levels or systematic effects. By Bayes' rule, our knowledge of these parameters, taking 
into account the data, is described by the posterior probability 
distribution \begin{eqnarray} p(S, L, \Phi | D) \propto 
p(S,L,\Phi)p(D|S,L,\Phi) \\ = p(S)p(L|S)p(\Phi|L,S)p(D|S,L,\Phi) 
\end{eqnarray} Different lens modelling techniques can all be 
interpreted within this framework \citep{2006ApJ...637..608B}. 
Differences between them are mostly due to the effective choices for the 
probability distributions $p(S,L,\Phi)$, describing prior knowledge 
about the system, and $p(D|S,L,\Phi)$, describing prior knowledge about 
how the observed data are related to the system. The other area where lens 
modelling techniques differ is the strategy for summarising the 
posterior distribution: typically one either searches for the peak to 
find a best-fitting model, or else one tries to obtain a sample of 
typical models from the posterior distribution. The latter approach is 
more computationally expensive but results in a more honest summary of 
the uncertainty in the conclusions, as the posterior probability of any hypothesis 
can be easily calculated. Throughout this paper, we will use a 
sampling-based approach, prioritising realism over convenience.

\subsection{A Realistic Prior on Sources}\label{priors} To begin the 
inference, we must construct a prior probability distribution to 
describe what kinds of sources are 
considered plausible. Ideally, the prior distribution 
should contain all of the relevant information that we have about 
typical light profiles of galaxies, but should otherwise make minimal 
assumptions about features of the light profile that we are actually 
uncertain about. We will now list features that an ideal prior distribution
for a galaxy source should have. Various priors have been proposed and used in the 
literature, usually satisfying some of these criteria but not
others. In Section~\ref{blobby} we develop a prior that has all of these 
properties. This is not to claim that our prior is optimal, or the best
one could do, merely that it avoids some of the most obvious flaws.

\begin{itemize} \item The surface brightness should be non-negative 
everywhere, with 100\% prior probability. This rules out the use of
``linear regularisation'', i.e. pixellated intensity maps with a joint Gaussian
prior for the intensity values \citep[e.g.][]{2006MNRAS.371..983S, 2008MNRAS.388..384D}. \\
\item The surface brightness may be close to zero over a 
large fraction of the domain \citep{2006ApJ...637..608B}. This rules out pixellated intensity maps with constrained uniform prior, and the ``maximum entropy'' prior \citep{wallington}. \\ \item 
Surface brightness at one point should be correlated with surface 
brightness at another nearby point. This correlation should decrease 
with distance.\\ \item The degree of spatial correlation may vary across 
the domain: i.e. we should not rule out the possibility of smoothness in 
one region and small scale structure in another. \\ \item The prior 
should be convenient to explore via Markov Chain Monte Carlo (MCMC). \\ 
\item There is probably some degree of correlation between the surface 
brightness profile in different wavebands, although this is not 
guaranteed. \end{itemize}

The statistics literature contains many proposals for general prior 
distributions for density functions (e. g. the Dirichlet distribution), which may be adapted for our 
purpose; however, the context here is different: our densities will be 
surface brightness profiles and not probability densities for data 
points. Thus, our concern is not so much with the mathematical 
properties of the priors but with which ones produce density profiles 
that resemble galaxies. We attempt to satisfy all of the design criteria 
listed above with our blobby prior described in the next section.

\subsubsection{Blobby Prior}\label{blobby} Our models of surface 
brightness profiles will ultimately be constructed from some basis 
functions, also known colloquially as ``blobs''. Specifically, we choose elliptical S{\'e}rsic profiles as our 
basis functions, since in some cases even a single S{\'e}rsic profile is 
sufficient to describe the light profile of a galaxy. This approach of 
using moderate numbers of physically motivated profiles is not new \citep[e.g.][]{2006MNRAS.372.1289M, massinf}, but is a simple way of implementing a prior with the desired properties outlined above.

In a suitably centred and aligned Cartesian coordinate system, a blob 
with scale radius $R_0$ and axis ratio $q$ has the following 
profile: \begin{equation}\label{sersic} f(x,y) = 
C\exp\left[-\left(\frac{R}{R_0}\right)^\alpha\right] 
\end{equation} where $R=\sqrt{x^2/q + qy^2}$. When $\alpha=2$, 
the blob is a Gaussian; $\alpha=1$ corresponds to an exponential 
profile, and $\alpha=1/4$ is the de Vaucouleurs profile. We restrict 
$\alpha$ to be between 0.1 and 10. We also allow the number of blobs, $\texttt{numBlobs}$, to be a free parameter.

Since we will be analysing multi-band images, we must also assign a 
prior probability distribution to the colours of the blobs. To accomplish this, each blob 
has a set of parameters called \texttt{unnormalisedSpectrum}, describing 
how bright the blob is in each band relative to the other bands. The 
blobs can all have similar colours, or the colours can vary immensely. 
The degree of diversity of the spectra is controlled by a hyperparameter 
called $\texttt{sigSpectrum}$. The hyperparameter $\texttt{meanSpectrum}$ describes the typical colour that the blobs are centred around.

The prior for the central positions of the blobs was chosen to be a
an independent Gaussian distribution for each blob, with a common
centre position and standard deviation, as hyperparameters. This allows
the blobs to be either close to concentric, or not, as required by the data.
If the blobs are almost concentric, then sums of several blobs can accurately
model most decreasing radial profiles. Multi-component sources can also be modelled,
they correspond to the case where the blobs are separated by more than about one
scale radius. We have experimented with more
sophisticated schemes to correlate the central positions of the blobs 
\citep{neal} but found that the gain in realism was minor, yet the 
computational complexity of the problem increased significantly.

The typical peak surface brightness of blobs was assigned a prior that spans several orders of magnitude, 
centred around unity. Thus, the reconstruction software expects images 
in units where the fluxes are of order close to unity. The actual peak surface brightness of each
blob has a lognormal prior, centered around the typical value.

\begin{figure*} \begin{center} \includegraphics[scale=0.75]{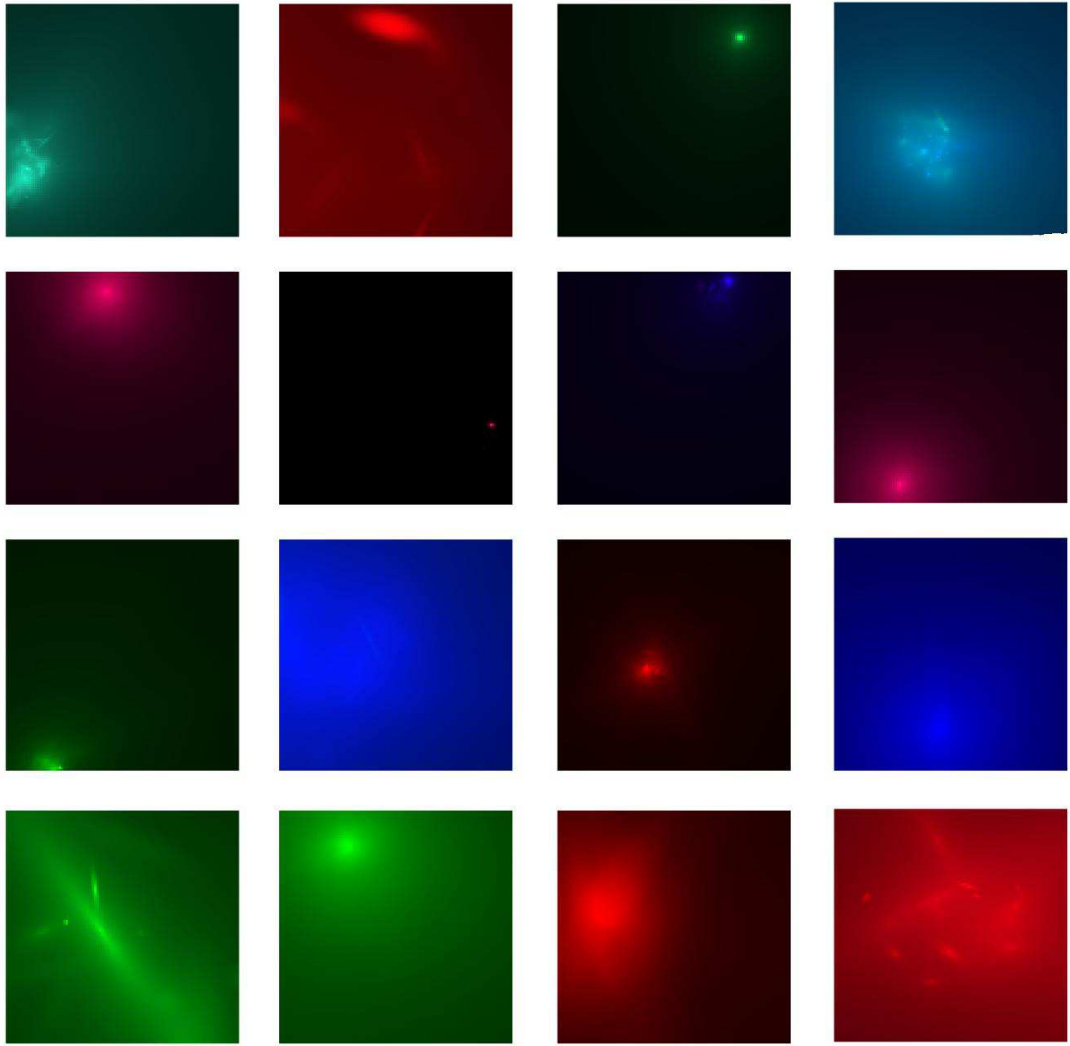} 
\caption{A sample of random light profiles (source galaxies generated from the prior 
distribution defined in Section~\ref{blobby}. Simple and complex 
profiles are possible, as are a variety of sizes and colours. The peak 
surface brightness is also uncertain by several orders of magnitude. 
While this prior clearly is not an optimal assignment, it is a more realistic assignment than  regularisation formulae \citep{2006MNRAS.371..983S}, for 
which the corresponding images would resemble noise (for first-order 
regularisation) or blurred noise (for gradient or curvature 
regularisation).\label{sample}} \end{center} \end{figure*}

For a thorough description of all of the 
parameters and the prior distributions that define our blobby prior, see Table~\ref{prior_table}.
Figure~\ref{sample} shows sample galaxy light profiles drawn from the prior. We recommend, as a general practice, generating
random models from the prior as a visual check on the reasonableness of the prior probability assignments.

\begin{table*} \begin{center} \caption{All of the parameters of a 
multi-wavelength blobby light profile, and their prior distributions. If 
the prior for a parameter refers to a hyperparameter, then this defines 
the prior conditional on that hyperparameter. The priors for the 
hyperparameters themselves are included in the second half of the table. All priors 
are independent except where otherwise specified. $J(a,b)$ denotes 
Jeffreys' scale-invariant prior $p(x) \propto 1/x$ between lower and 
upper bounds of $a$ and $b$ respectively. Lognormal$(a,b^2)$ denotes
a lognormal distribution, with $a$ being the central value for the variable, and $b$ being the standard deviation for the log of the variable. Note that $a$ may be named $\texttt{mean}*$, even though it is not technically the mean of the distribution.\label{prior_table}} 

\begin{tabular}{lll} \hline {\bf Parameter} & {\bf Definition} & {\bf 
Prior}\\ \hline Individual Blob Parameters & & \\
\hline $\texttt{peak}$ 
& Peak surface brightness, $C$ in Eqn~\ref{sersic} & 
Lognormal(\texttt{meanPeak}, \texttt{stdevLogPeak}$^2$) \\ 
$\texttt{width}$ & Scale width ($R_0$ in Eqn~\ref{sersic}) & Exponential(\texttt{meanWidth})\\ 
$\texttt{xc, yc}$ & Central position & Normal(($X_c, Y_c$), $\texttt{spread}^2$)\\ 
$\texttt{q}$ & Axis ratio & Uniform(0.1,1)\\ $\texttt{theta}$ & angle of 
Orientation & Uniform$(0, 2\pi)$\\ $\texttt{slope}$ & Slope of light 
profile, $\alpha$ in Eqn~\ref{sersic}& Lognormal($\texttt{meanAlpha}, \texttt{sigAlpha}^2$)\\ 
$\texttt{unnormalisedSpectrum}$ & Share of bolometric flux at each 
wavelength & Lognormal$(\texttt{meanSpectrum}, \texttt{sigSpectrum}^2)$\\ \hline 
Hyperparameters & & \\ \hline $\texttt{numBlobs}$ & Number of blobs & 
Uniform($\left\{0, 1, 2, ..., 10\right\}$) \\ 
$(X_c, Y_c)$ & Typical position that blobs are centred around & Uniform over source domain \\
$\texttt{spread}$ & Standard deviation of blob positions & $J(10^{-3}, 1) 
\times$ (image width) \\
$\texttt{meanPeak}$ & 
Typical brightness of a blob & $J(10^{-3}, 10^3)$ \\ 
$\texttt{stdevLogPeak}$ & Diversity of brightness of blobs & $J(0.03, 2)$ 
\\ $\texttt{sigSpectrum}$ & Diversity of spectra & $J(0.01, 3)$ \\ 
$\texttt{meanWidth}$ & Typical scale width of a blob & $J(10^{-3}, 1) 
\times$ (image width)\\ 
$\texttt{meanSpectrum}$ & Typical spectrum that blob spectra are centred around & Lognormal(0, $3^2$)\\
$\texttt{meanAlpha}$ & Typical slope & $J(0.1, 10)$ \\
$\texttt{sigAlpha}$ & Diversity of slopes & $J(0.01, 2)$ \\
\hline\end{tabular}
\medskip\\ \end{center} 
\end{table*}

\subsection{Lens Models}\label{lens} In cases where the light profile of the lens
galaxy appears simple, we assume that the mass profile is also simple and model the projected mass profile 
of the lens galaxy as a nonsingular isothermal ellipsoid (NIE) plus external 
shear ($\gamma$). This model (and related models, e.g. singular isothermal ellipsoid without external shear) has been used very successfully to model 
the mass profiles of elliptical galaxy lenses \citep[e.g.][]{slacs}; 
even when more general models have been used, the mass profiles are 
inferred to be close to isothermal. The model has eight parameters: $b$, 
the ``Einstein radius'', $q$, the axis ratio of the major and minor axes of 
the elliptical mass distribution, $(x_c, y_c)$, the central position of 
the lens, $\theta$, the orientation angle of the lens, $\gamma$, the 
strength of the external shear, $\theta_\gamma$, the orientation angle 
of the external shear, and $r_c$, the core radius (often set to zero, giving a singular isothermal ellipsoid). A simple interpretation
of $b$ is the length of the minor axis of the critical curve, in the singular case $r_c = 0$.

The source plane position $(x_s, y_s)$ and the image plane position 
$(x,y)$ of a ray are related by: \begin{eqnarray}\label{lenseqn} x_s = x 
- \alpha_x(x,y) \nonumber \\ y_s = y - \alpha_y(x,y) \end{eqnarray} 
where the deflection angles $\alpha$ are given by 
\citet{1998ApJ...495..157K}. Defining $\psi = \sqrt{q^2(r_c^2 + x^2) + 
y^2}$, the deflection angle formulae are: \begin{eqnarray} \alpha_x(x,y) 
&=& \frac{b}{\sqrt{1-q^2}}\tan^{-1}\left[\frac{x\sqrt{1-q^2}}{\psi + 
r_c}\right]\\ \alpha_y(x,y) &=& 
\frac{b}{\sqrt{1-q^2}}\tanh^{-1}\left[\frac{y\sqrt{1-q^2}}{\psi + q^2 
r_c}\right] \end{eqnarray} as long as $q<1$. If $q>1$, then $q$ can 
simply be replaced by $q^{-1}$ and the orientation angle $\theta$ 
rotated by 90 degrees, with the same result. We do not allow $q=1$ but 
it can become arbitrarily close to 1 and therefore close-to-spherical 
lenses are not ruled out. By making the 
prior for the lens strength $b$ depend on the image size, we are able to 
avoid the need for system-specific tuning of the priors: we simply 
assume that the input images are such that the Einstein Radius $b$ of the 
lens is of the same scale as the input image.

\subsection{Complex Lenses}\label{complex} Since the SLUGS sample 
contains very massive lenses, many of the lenses are not smooth galaxies 
but are instead galaxy groups. For modelling these systems, the NIE family is not flexible enough to represent the 
complex structure of the lens. To overcome this, it is possible to use the blobby prior 
of Section~\ref{blobby}, but interpret each blob as a nonsingular
isothermal ellipsoid (with zero external shear) rather than a S{\'e}rsic 
profile. This is straightforward as there is a simple correspondence 
between the parameters of the S{\'e}rsic profile and the parameters of 
the NIE. However, this creates computational inefficiencies, as the 
code needs to infer the positions and orientations of many lens components, and the likelihood function tends to be more difficult when modifying the lens as opposed to the source. We note that gravitational interactions within groups will tend to make the NIE approximation for each galaxy less valid, however sums of NIEs remain a tractable and flexible family of models.

For complex lenses in the SLUGS sample, we use the following strategy. First, we visually identify the approximate positions, ellipticities and orientations of the dominant lens components. The prior for these parameters can then be set to be narrow Gaussians centred on the estimated values. The Einstein radii $\{b\}$
of these lenses can then be assigned a broad prior, such as independent $\propto 1/b$ priors within a generous
range. This prior suggests that from the images, we know where the lens substructures are, but we do not know their relative masses, because the mass to light ratio might differ between objects.

\section{Sampling Distributions}\label{sampling} The prior distributions 
for the unknown parameters have now all been defined. However, we must 
also assign sampling distributions $p(S,L,\Phi| D)$ for all $D$ and 
$(S,L,\Phi)$ (these are known as sampling distributions when $D$ is 
unknown, when $D$ is fixed at the observed data, $p(S,L,\Phi| D)$ as a 
function of the parameters is called the likelihood function). For any 
value of the source and lens parameters $(S,L)$ we can compute mock 
images $M(S,L)$ by ray tracing through the lens (firing 4 rays per image 
pixel, a compromise value chosen for efficiency) and blurring by the PSF (assumed known from nearby stars in the 
field). The likelihood is then related to the $\chi^2$ deviation between 
the mock image and the actual image; i.e. we assume a sampling 
distribution for the noise that is normal/Gaussian with mean 0, and independent 
for each pixel: \begin{eqnarray} p(D|S,L,\Phi) = \nonumber \\ 
\exp\left[-\frac{1}{2}\sum_{i=1}^n \left\{\log(2\pi\sigma_i^2) + 
\left(\frac{D_i - M(S,L)_i}{\sigma_i}\right)^2\right\}\right] 
\end{eqnarray} where $n$ is the total number of pixels, and the standard 
deviation for each pixel is given by: \begin{equation} \sigma_i = 
\sqrt{\sigma_{{\rm map}, i}^2 + \sigma_{{\rm base}}^2 + \left(\tau 
\sqrt{M(S,L)_i}\right)^2} \end{equation} Here, $\sigma_{{\rm map}}$ is the 
uncertainty reported from the weight map, and the extra parameters 
$\sigma_{{\rm base}}$ and $\tau$ allow for inaccuracies in the weight 
map, by possibly boosting the noise level by a constant value, or a 
value proportional to the square root of the predicted surface brightness. 
$\sigma_{{\rm base}}$ and $\tau$ were assigned $J(10^{-3}, 10^3)$ 
priors. Once again, this assumes that the input images are in units 
where the typical pixel values are of order unity.

Although it is not presented in this paper, we also recommend generating
simulated data sets from the sampling distribution, using models drawn from
the prior. This practice gives visual checks on the {\it joint prior}, $p(\theta,D)=p(\theta)p(D|\theta)$
which inference depends on.

The computational task of using images to infer appropriate values for 
all of the above parameters, with uncertainties, is rather difficult. To 
accomplish this goal, we use a variant of Nested Sampling 
\citep{skilling}, designed to cope effectively with multimodal and 
highly correlated posterior distributions. This algorithm is described briefly in
Section~\ref{nested_section}. For a full description, see \citet{dnest}.

\section{Nested Sampling Implementation}\label{nested_section} Nested 
Sampling (NS) is a powerful and widely applicable algorithm for Bayesian 
computation \citep{skilling}. Starting with a population of particles 
$\{\theta_i\}$ drawn from the prior distribution $\pi(\theta)$, the 
worst particle (lowest likelihood $L(\theta)$) is recorded and then 
replaced with a new sample from the prior distribution, subject to the 
constraint that its likelihood must be higher than that of the point it 
is replacing. As this process is repeated, the population of points 
moves progressively higher in likelihood.

Each time the worst particle is recorded, it is assigned a value $X \in 
[0,1]$, which represents the amount of prior mass estimated to lie at a 
higher likelihood than that of the discarded point. Assigning $X$-values 
to points creates a mapping from the parameter space to $[0,1]$, where 
the prior becomes a uniform distribution over $[0,1]$ and the likelihood 
function is a decreasing function of $X$. Then, the evidence can be 
computed by simple numerical integration, and posterior weights can be 
assigned by assigning a width to each point, such that the posterior 
mass associated with the point is proportional to \texttt{width} 
$\times$ \texttt{likelihood}.

The key challenge in implementing Nested Sampling for real problems is 
to be able to generate the new particle from the prior, subject to the 
hard likelihood constraint. If the discarded point has likelihood $L^*$, 
the newly generated point should be sampled from the constrained 
distribution: \begin{equation}\label{constrained} p_{L^*}(\theta) = 
\frac{\pi(\theta)}{X^*}\left\{\begin{array}{lr}1, & L(\theta) > L^*\\ 0, 
& \textnormal{otherwise.}\end{array}\right. \end{equation} where $X^*$ 
is the normalising constant. Technically, our knowledge of this new 
point should be independent of all of the surviving points. A simple way 
to generate such a point is suggested by \citet{sivia}: copy one of the 
surviving points and evolve it via MCMC with respect to the prior 
distribution, rejecting proposals that would take the likelihood below 
the current cutoff value $L^*$. This evolves the particle with 
Equation~\ref{constrained} as the target distribution. If the MCMC is 
done for long enough, the new point will be effectively independent of 
the surviving population and will be distributed according to 
Equation~\ref{constrained}. However, in complex problems, this approach 
can easily fail - constrained distributions can often be very difficult 
to efficiently explore via MCMC, particularly if the posterior 
distribution is multimodal or highly correlated. To overcome these 
drawbacks, sophisticated schemes such as MultiNest 
\citep{2008arXiv0809.3437F} have been developed and applied successfully 
in low-dimensional problems. However, for our purposes we require a
method that uses MCMC.

The main advantage of Nested Sampling is that successive constrained 
distributions (i.e. $p_{L^*_j}(\theta), p_{L^*_{j+1}}(\theta)$, and so 
on) are, by construction, all compressed by the same factor relative to 
their predecessor. This is not the case with tempered distributions of 
the form $p_T(\theta) \propto \pi(\theta)L(\theta)^{1/T}$, where a small 
change in temperature $T$ can correspond to a small or a large 
compression. Tempering based algorithms (e.g. simulated annealing, 
parallel tempering) will fail unless the density of temperature levels 
is adjusted according to the specific heat, which becomes difficult at a 
first-order phase transition. 
Unfortunately, knowing the appropriate values for the temperature levels 
is equivalent to having already solved the problem! Nested Sampling does 
not suffer from this issue because it asks the question ``what should 
the next distribution be, such that it will be compressed by the desired 
factor'', rather than the tempering question ``the next distribution is 
pre-defined, how compressed is it relative to the current 
distribution?''

\subsection{Diffusive Nested Sampling}\label{linked}
\begin{figure*}
\includegraphics[scale=0.9]{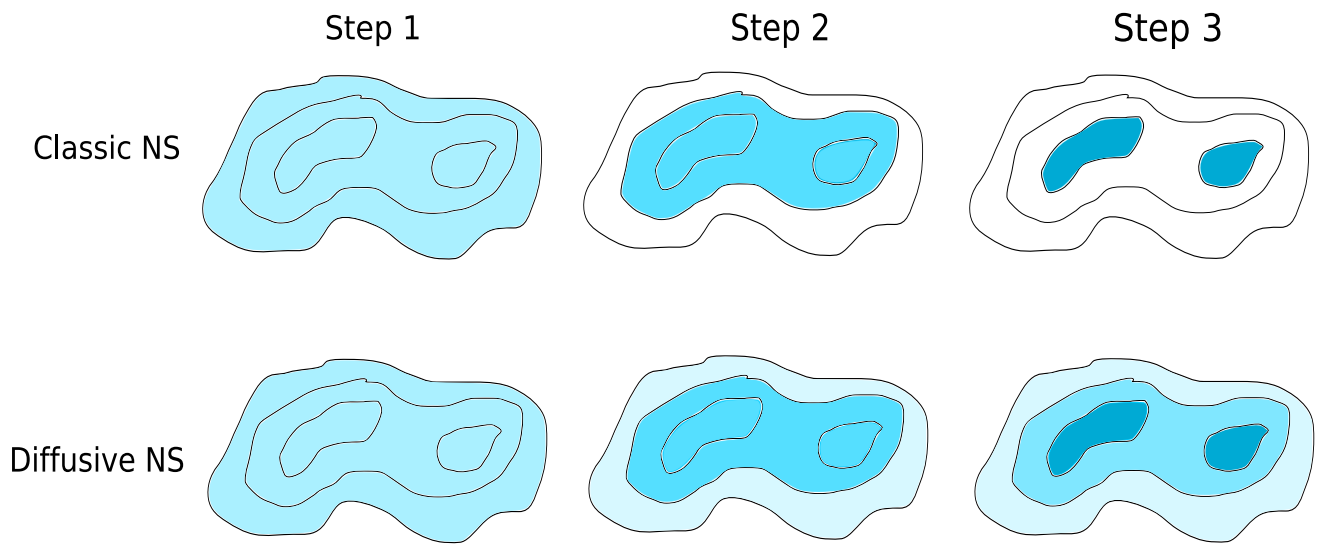}\caption{An illustration of a the distributions that must be sampled as Nested Sampling progresses \citep{dnest}. In the classic scheme, at Step 3, we must obtain a sample from the coloured region which is composed of two separate islands, which is usually very difficult if MCMC is the only exploration option. To overcome these difficulties, we explore the mixture distribution (bottom right), where travel between isolated modes is more likely.\label{nested}}
\end{figure*}

Throughout this paper we use the efficient Diffusive Nested Sampling method
\citep{dnest} which has been found to speed up MCMC-based Nested Sampling
by an order of magnitude. The basic idea is that, instead of exploring
a single constrained distribution, the particle explores a mixture of
the required constrained distribution, and the past constrained distributions
(See Figure~\ref{nested}).
New likelihood contours are created using likelihoods accumulated during
the run, which provides more information than using only the likelihood
value of the endpoint of an MCMC run. Finally, the absence of the copying
operation prevents the depletion problem that can occur in the classic
implementation of NS, where repeated copying destroys the diversity of the
population of particles prematurely. For more details, see \citet{dnest}.

\subsection{Nested Sampling Solves Multiple Problems at Once} Nested 
Sampling algorithms provide samples from the parameter space, along with 
an assignment of prior mass (width in the $X$-space) to each sample. 
These samples can be used to probe (generate samples and estimate the 
normalising constant) target probability distributions other than the 
posterior. For example, a common family of distributions of interest are 
the tempered posterior distributions: \begin{equation} p_T(\theta) = 
\frac{1}{Z(T)} \pi(\theta) L(\theta)^{1/T} \end{equation} 
These distributions can be studied by weighting the particles according
to $\left(\texttt{prior mass}\right)$ $\times L^{1/T}$, for various $T$ rather than just
$T = 1$.

Of course, 
this is only possible if the sample contains points from regions of 
parameter space that are significant with respect to the modified target 
distribution.

This technique can be used to solve the inference problem under 
different prior probability assignments at very little computational 
cost. Particularly, we can explore the effect of different sampling 
distributions (Section~\ref{sampling}). For example, suppose we 
hypothesise a correlated probability distribution for the noise. This 
means that larger, correlated deviations between the model and the data 
can be ascribed to noise, rather than model inadequacy. Thus, samples 
with lower $X$-values will be deemed acceptable and included in the 
sample. The effect of correlated noise models is very similar to the 
effect of simply raising the temperature, an operation that is 
computationally trivial.

Note that not all systematic errors can be accounted for in this way. 
For example, imagine that the sky background had been subtracted 
incorrectly, leaving a gradient across the image. This would be modelled 
using light in the source plane. If we attempted to reweight the sample 
by using the marginal likelihood after integrating over a prior 
distribution for the incorrect sky subtraction, we would find that one 
point dominated the sample. This immediately provides evidence of 
failure: essentially, Nested Sampling would have scaled the wrong peak in parameter 
space, and did not obtain any significant samples with respect to the 
modified posterior. This situation is easily identified (only a single particle will have significant weight) and is analogous to using a poor approximating 
distribution in standard importance sampling.

\section{CSWA 31}\label{cswa31} 
\begin{figure*}
\begin{center} 
\includegraphics[scale=0.7]{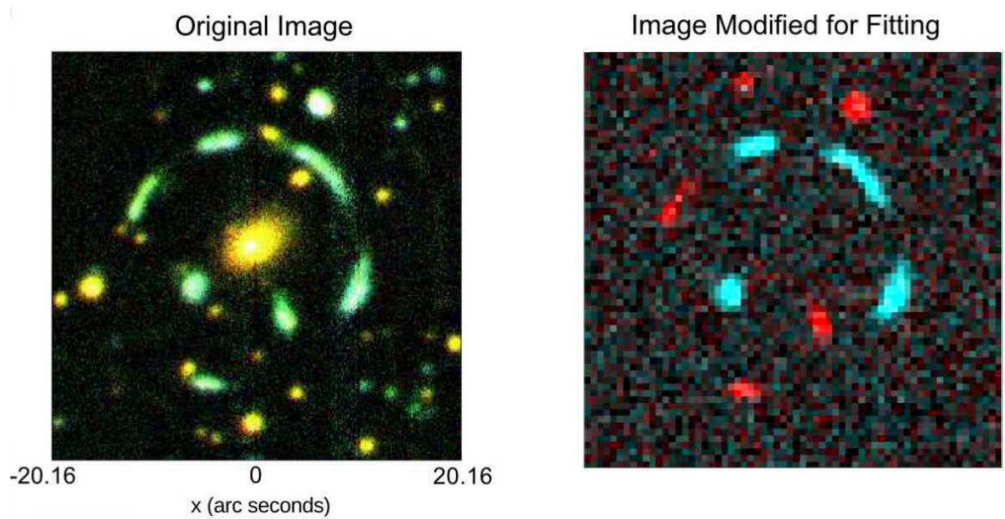} \caption{SLUGS Gemini/GMOS 
Image of CSWA 31, with the primary lens galaxy subtracted (left). The 
right hand image shows the bluest image of CSWA 31 (where the redder lens 
substructures are not visible), resampled onto a 64$\times$64 pixel 
grid, for computational speed. Only the brightest lensed images were retained. Preliminary studies indicated that four of the images (the blue images in the modified image) are a classic quad configuration. For computational reasons we gave these images a common colour, to help our algorithm find the correct model more quickly.
\label{cswa31_full}}
\end{center}
\end{figure*}

CSWA 31 is a complex lens with at least seven 
identifiable images and a very large image 
separation ($\sim$ 12'')(Figure~\ref{cswa31_full}). The lens is a galaxy group, with a dominant 
central galaxy acting as the primary lens. We obtained g,r, and i-band images for CSWA31 on February 21, 2009, using the GMOS instrument on the Gemini South telescope. 1200 second dithered exposures were obtained in each of the bands, in seeing ranging from 0.7-0.75''. The GMOS field of view is 5.5 x 5.5 arcmin, and with 2x2 binning, the image scale is 0.145''/pixel. In Figures 3-5, north is towards the bottom of the page.

\begin{figure*}
\begin{center}
\includegraphics[scale=0.5]{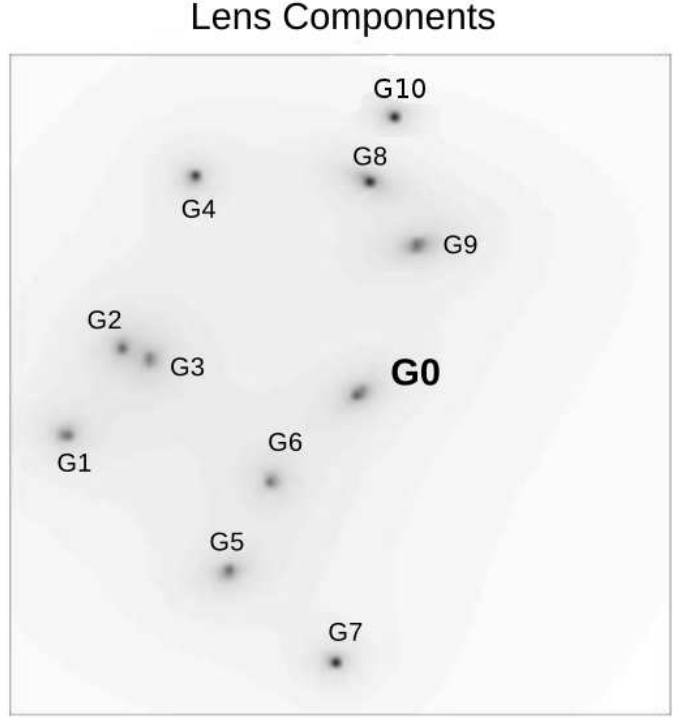}
\caption{Mean positions of the SIE lenses for the complex lens model for CSWA 31. The prior uncertainty on the positions is 1''. The relative masses are left as free parameters, with broad priors. The eleven lens components chosen were the ones that appear most luminous and close to images, and therefore most likely to be significant lenses. Not all of the yellow components in Figure~\ref{cswa31_full} were included for computational efficiency. \label{complex_lens}}
\end{center}
\end{figure*}

Due to the presence 
of mass substructure (as seen by the large number of yellow blobs in the image), we used the complex lens model 
(Section~\ref{complex}) for this system. The dominant yellow sources in the image
were modelled as 11 SIE lenses, with their positions, ellipticities and orientation
angles almost-fixed at the observed values from the light (the prior allowed for
a small uncertainty in each). We allowed the Einstein radii $\{b\}$ to be free parameters
with independent scale-free priors, truncated such that the primary lens must
have the highest $b$. The central positions of the lens blobs are shown in Figure~\ref{complex_lens}.

\begin{figure*}
\begin{center}
\includegraphics[scale=0.6]{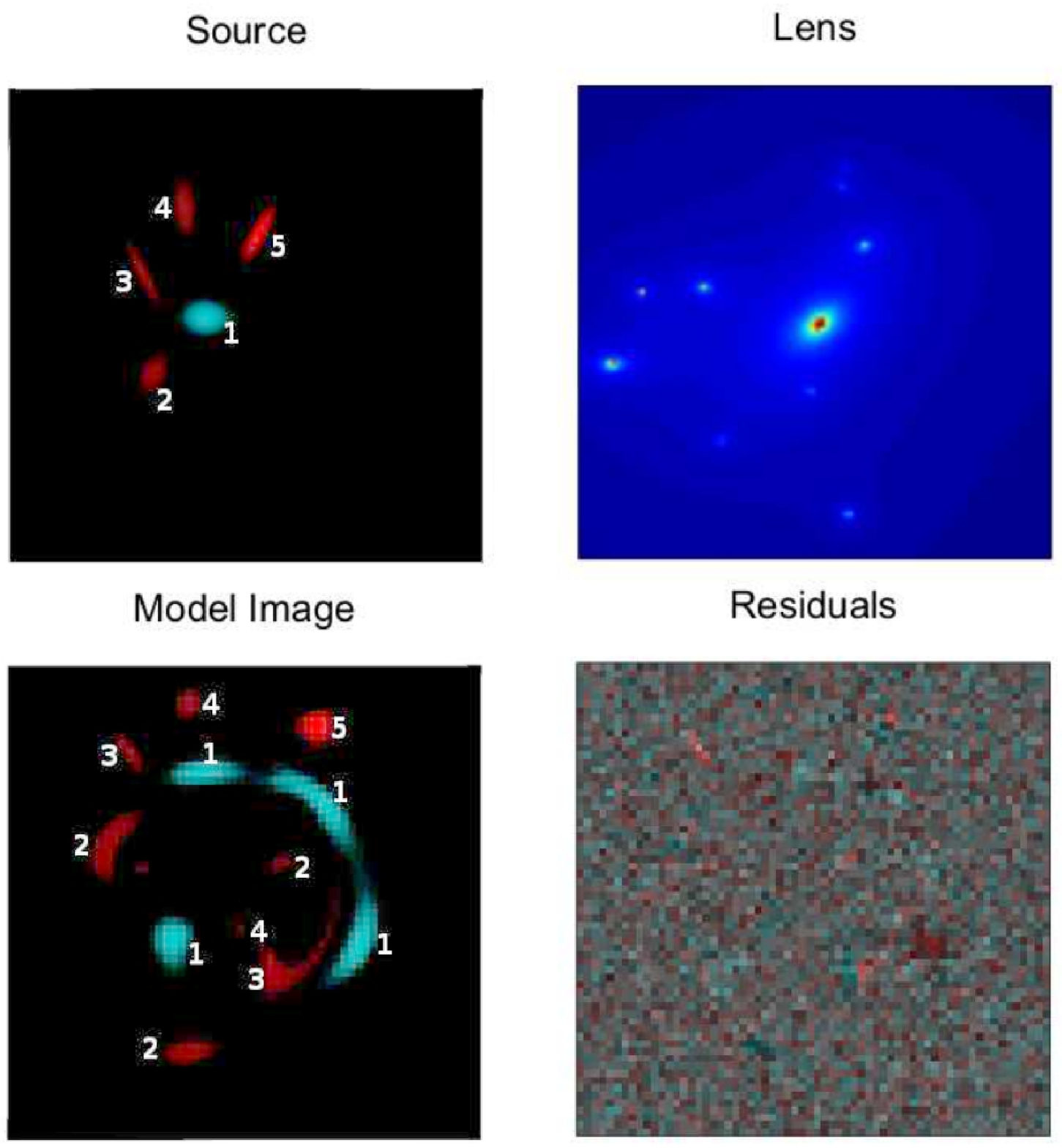}
\caption{A typical model from the posterior distribution for the CSWA 31 reconstruction. The source is composed of five components, separated on the sky by $\sim$ 3 arcseconds. The primary source (1) is lensed into a classic quad configuration. (2) has been lensed in a complex manner into three main images, and (3) is doubly imaged,\label{cswa31_result}}
\end{center}
\end{figure*}

Unfortunately the complex lens model creates 
large computational overheads: the time taken to simulate an image 
increases with the number of lens ``blobs''. Additional computational 
expense is caused by the fact that the parameter space now has more 
dimensions, causing Diffusive Nested Sampling to require more steps per 
new level, {\it and} more levels to reach the bulk of the posterior. 
Finally, the large separation of the lens implies that the image simply 
has more pixels. To reduce the effect of these computational challenges, 
we analyse only a degraded resolution version of the CSWA 31 image, containing only the most significant images 
(Figure~\ref{cswa31_full} shows the full and the degraded image). This means our final reconstruction will have only
the brightest of the source components that are actually present.

A further challenge is caused by the fact that it is not clear which
sources are in the lens plane and which are background sources, and
even whether the background sources are all at the same redshift. For
this initial study, we assume that all images in the right panel of Figure~\ref{cswa31_full}
are images of sources lying at a single redshift.

The modelling was run with a maximum of five sources. A typical model from the run
is shown in Figure~\ref{cswa31_result}. The source is comprised of five galaxies
of extent $\sim 2''$, separated by $\sim 5''$.  The source (1) has been lensed into
a classic quad configuration by the primary lens. Note that, according to our model,
the leftmost image of (2) and the rightmost image of (3) are different sources and
so may be different colours,
whereas they appear to match in the image (Figure~\ref{cswa31_full}). We attempted
to find models where these two images are images of the same source, but we were
unable to do so. The image in Figure~\ref{cswa31_full} shows additional small green
sources that may also lie scattered throughout the source group.

Unfortunately the source redshifts are
unknown so the physical scale is also unknown, however we are able to provide an estimate.
The redshift of the lens
is 0.683. Assuming a source redshift $z_s=1$ gives a source plane scale of
8.3 kpc$/''$, implying that these source galaxies are large spirals. Assuming $z_s=0.8$ gives 7.7 kpc$/''$ and $z_s=2$ gives 9.3 kpc$/''$, so the physical interpretation is insensitive to
the source redshift. The magnification of the source is $2.8 \pm 0.1$ magnitudes, a factor
of $\sim 14$.

We measure the mass of the primary lens, within its critical curve, to be $1.4 \pm 0.2 \times 10^{13} M_{\odot}$. The other lenses masses are significantly uncertain, so we do not report them here.

\section{Conclusions}\label{conclusions} In this paper we have detailed 
the lens modelling technique to be used for the complex gravitational 
lenses in the SLUGS (Sloan Lenses Unravelled by Gemini Studies) sample. 
The main features of our technique are the attempt to incorporate 
realistic prior information about galaxy surface brightness profiles, 
and the Diffusive Nested Sampling method for exploring the posterior 
distribution for the reconstructions. We recommend drawing random samples
from complex prior distributions as a visual check on the reasonableness
of the probability assignments. Our random samples qualitatively resemble
simple galaxy profiles, whereas the corresponding random samples from
other methods, such as linear regularisation, are simply noise.
Hence, our method uses much more prior information and can be expected to perform better than the alternatives when the data are noisy or low resolution.

We presented the first reconstruction of the 
complex SLUGS lens CSWA 31. We found that the observed lensing configuration is best explained by five source components, lensed into the elaborate image structure by the complex group lens. Although the source redshift is unknown, it is plausible that the physical sizes of the sources are $\sim 10$ kpc and hence we are observing
the merger of five large spiral galaxies.

\section{Acknowledgements} BJB is grateful to Lianne Zimmermann for helping him to settle in 
to Santa Barbara. I would also like to thank Matt Auger and Phil Marshall
for advice about modelling CSWA 31, and Tommaso Treu for valuable comments.

\end{document}